\documentstyle[preprint,aps,epsf,tighten]{revtex}	
\def\aprle{\buildrel < \over {_{\sim}}}

\begin{document}
%
%
\preprint{$
\begin{array}{l}
\mbox{FERMILAB-Pub-04/114-T}\\
\mbox{hep-ph/0407155}\\
\mbox{August 2004}\\[0.5in]
\end{array}
$}
\title{Normal vs. Inverted Hierarchy in Type I Seesaw Models\\}
\author{Carl H. Albright}
\address{Department of Physics, Northern Illinois University, DeKalb, IL
60115\\
       and\\
Fermi National Accelerator Laboratory, P.O. Box 500, Batavia, IL
60510\footnote{electronic address: albright@fnal.gov}}
\maketitle
\begin{abstract}

We demonstrate that, for every grand unified model based on a conventional
type I seesaw mechanism leading to a normal light neutrino mass hierarchy, 
one can easily generate a corresponding model with an inverted hierarchy which 
yields the same neutrino oscillation parameters.  However, the latter type model 
has several unattractive instabilities which will disfavor any grand unified 
type~I seesaw model, if an inverted neutrino mass hierarchy is observed 
experimentally.  This should be contrasted  with the softly-broken  
$L_e - L_\mu - L_\tau$ flavor symmetry models which are eliminated, if the data 
favors a normal mass hierarchy.
\thispagestyle{empty}
\end{abstract}
\vskip 2in
PACS numbers: 14.60.Pq, 12.10.Dm, 12.15Ff, 12.60.Jv\\
\newpage
%

Since the first convincing evidence in 1998 for atmospheric neutrino 
oscillations by the Super-Kamiokande (S-K) collaboration and 
subsequent confirmation of hints of solar neutrino oscillations 
by both the Sudbury Neutrino Observatory and KamLAND collaborations, great 
progress has been made in pinning down the values of the leading oscillation 
parameters $\Delta m^2_{32},\ \sin^2 2\theta_{23},\ \Delta m^2_{21}$ and 
$\sin^2 2\theta_{12}$ \cite{data}.  The same can not be said for 
the other 3-family mixing parameters, $\sin^2 2\theta_{13}$ and the leptonic
Dirac and Majorana CP phases $\delta,\ \chi_1$ and $\chi_2$.  The angle 
$\theta_{13}$ has been bounded by the CHOOZ and Palo Verde reactor experiments
\cite{reactor} to be noticeably smaller than the other two angles, while the three
phases are completely unknown.  Moreover, most important for the subject of
this letter, even the 3-neutrino mass hierarchy, normal or inverted, 
remains undetermined.  Experimental determination of the hierarchy and 
$\theta_{13}$ will be critical for the continued viability of models 
in the literature which attempt to explain the neutrino mixing data; cf. 
\cite{rev} for extensive reviews of such models.

For models based on some flavor symmetry, those exhibiting a softly broken
$L_e - L_\mu - L_\tau$ lepton number produce only an inverted hierarchy 
\cite{inverted}.  In contrast, all of the successful grand unified models in 
the literature employing a conventional type I seesaw mechanism appear to 
prefer a normal hierarchy; cf. \cite{normal} for some still viable models of
this type.  We seek to explore here why this is the case.  With such a dichotomy 
present, clearly identification of the correct hierarchy observed in Nature will 
serve to eliminate a large batch of models of one kind or the other.  On the 
other hand, grand unified models based on a type II \cite{II} or type III 
\cite{III} seesaw with direct or induced Higgs triplet contributions can be 
constructed by design to have either a normal or an inverted hierarchy.

In this letter we show how, given a satisfactory grand unified model with 
normal neutrino mass hierarchy based on a type I seesaw, one can easily 
generate a corresponding model with an inverted hierarchy which also
satisfies the same known neutrino oscillation parameters.  The 
issue becomes which hierarchy model is more satisfactory theoretically in terms 
of flavor symmetry, heavy neutrino mass hierarchy with possibly successful 
leptogenesis, and stability of the solution.  To address this
point, we use a much studied $SO(10)$ grand unified model with a $U(1) \times Z_2 
\times Z_2$ flavor symmetry proposed by the author in collaboration with S.M. 
Barr which leads naturally to a normal hierarchy \cite{ab}.  Results for the 
light neutrino mass matrices are also derived for the class of models in
which the charged lepton mass matrix is diagonal in flavor space.  Features 
obtained in the models considered are more generally expected to hold for the 
whole class of grand unified models with a type~I seesaw mechanism.

We begin with the known neutrino oscillation data and the definitions that
apply for the parameters appropriate for a normal vs. inverted hierarchy.
Experimentally the atmospheric and solar neutrino mixing parameters are 
approximately equal to \cite{data,reactor}
\begin{equation}
\begin{array}{rclrcl}
	|\Delta m^2_{32}| &\simeq& 2.5 \times 10^{-3}\ {\rm eV^2}, \quad & 
		\sin^2 2\theta_{23} &\simeq& 1.0, \\[0.1in]
	\Delta m^2_{21} &\simeq& 7.0 \times 10^{-5}\ {\rm ev^2}, \quad & 
		\tan^2 \theta_{12} &\simeq& 0.40, \\[0.1in]
	\Delta m^2_{31} &\simeq& \Delta m^2_{32}, \quad &
		\sin^2 2\theta_{13} &\aprle& 0.16. 
\label{eq:mix}
\end{array}
\end{equation}
Although the mass squared differences are determined, the actual neutrino
mass scale is undetermined as is the mass hierarchy, i.e., whether 
$\Delta m^2_{32} \equiv m^2_3 - m^2_2$ is positive or negative.  Clearly, 
however, the larger (smaller) mass separation is associated with the 
atmospheric (solar) neutrino flavor mixing.  For the solar neutrinos, 
the mass squared difference $\Delta m^2_{21}$ is known to be positive, since
the solar mixing angle $\theta_{12}$ lies in the first octant.

The Maki-Nakagawa-Sakata (MNS) neutrino mixing matrix with elements 
$U_{\alpha i}$ relates the flavor state $\alpha = e,\ \mu,\ \tau$ to the 
mass eigenstates $i = 1,\ 2,\ 3$ in their respective orders.  By convention,
we have 
\begin{equation}
	U_{MNS} = \left(\matrix{c_{12}c_{13} & s_{12}c_{13} & s_{13}e^{-i\delta}\cr
		-s_{12}c_{23} - c_{12}s_{23}s_{13}e^{i\delta} & 
		c_{12}c_{23} - s_{12}s_{23}s_{13}e^{i\delta} & s_{23}c_{13}\cr 
		s_{23}s_{12} - c_{12}c_{23}s_{13}e^{i\delta} & 
	     - c_{12}s_{23} - s_{12}c_{23}s_{13}e^{i\delta} & c_{23}c_{13}\cr}
		\right),
\label{eq:MNS}
\end{equation}
where the cosines and sines refer to rotations of the mass eigenstates.
From the mixing data in Eq. (\ref{eq:mix}), we can approximate $c_{23} \simeq
s_{23} \simeq 1/\sqrt{2}$, $s_{12} \simeq \sqrt{2/7},\ c_{12} \simeq 
\sqrt{5/7}$, and $s_{13} \aprle 0.2$.  The mass eigenstates are then 
approximately given in terms of the flavor eigenstates by 
\begin{equation}
\begin{array}{rcl}
	\nu_1 &\simeq& \sqrt{\frac{5}{7}}c_{13}\nu_e - \sqrt{\frac{1}{7}}
		(\nu_\mu - \nu_\tau) - \sqrt{\frac{5}{14}}s_{13}e^{-i\delta}
		(\nu_\mu + \nu_\tau), \\[0.1in]
	\nu_2 &\simeq& \sqrt{\frac{2}{7}}c_{13}\nu_e + \sqrt{\frac{5}{14}}
		(\nu_\mu - \nu_\tau) - \sqrt{\frac{1}{7}}s_{13}e^{-i\delta}
		(\nu_\mu + \nu_\tau), \\[0.1in]
	\nu_3 &\simeq& s_{13}e^{i\delta}\nu_e + \sqrt{\frac{1}{2}}c_{13}(\nu_\mu 
		+ \nu_\tau),
\label{eq:massstates}
\end{array}
\end{equation}
where the states $\nu_1$ and $\nu_2$ are involved with the solar neutrino
transitions, while $\nu_2$ and $\nu_3$ are involved with the atmospheric
neutrino transitions.  As such, the same $U_{MNS}$ mixing matrix 
leading to the same mass eigenstate flavor compositions applies, whether the 
hierarchy is normal, $m_1 < m_2 < m_3$, or inverted, $m'_3 < m'_1 < m'_2$,
as illustrated in Fig. 1.  For the inverted case, $\Delta m^2_{21} = 
m'^2_2 - m'^2_1$ and $|\Delta m^2_{32}| = m'^2_2 - m'^2_3$ with the mass
square differences specified experimentally in Eq. (\ref{eq:mix}).

We now turn to the issue of inverted vs. normal mass hierarchy in grand unified
models with the conventional type I seesaw mechanism.  Since most, if not all,
such models in the literature \cite{normal} favor a normal hierarchy, we start 
with that case.  The specified Dirac and right-handed Majorana mass matrices, 
$M_N$ and $M_R$, will both have normal hierarchy in that their mass eigenvalues 
with the smaller values will be more closely spaced.  The light left-handed 
neutrino mass matrix follows from the conventional type I seesaw formula 
\cite{gmrsy} according to 
\begin{equation}
	M_\nu = - M_N M^{-1}_R M^T_N,
\label{eq:seesaw}
\end{equation}
which by assumption also has a normal hierarchy, albeit a rather mild one.
This mass matrix is diagonalized by the unitary transformation,
$U_{\nu_L}$, whereby
\begin{equation}
	U^\dagger_{\nu_L} M^\dagger_\nu M_\nu U_{\nu_L} = {\rm diag}(m^2_1,\ m^2_2,
		\ m^2_3).
\end{equation}
Since the $M_\nu$ matrix is complex symmetric, one can choose phases for 
$U_{\nu_L}$ such that diagonalization of $M_\nu$ itself occurs with 
\begin{equation}
	U^T_{\nu_L} M_\nu U_{\nu_L} = {\rm diag}(m_1,\ m_2,\ m_3),
\label{eq:mnudiag}
\end{equation}
where the mass eigenvalues are real and positive with $m_1 < m_2 < m_3$.
The MNS mixing matrix is given by the product of the two unitary matrices,
$U_{L_L}$ which diagonalizes the charged lepton mass matrix, and $U_{\nu_L}$
above according to  
\begin{equation}
\begin{array}{rcl}
	V_{MNS} &\equiv& U^\dagger_{L_L}U_{\nu_L} = U_{MNS}\Phi,\\[0.1in]
		\Phi &=& {\rm diag}(e^{i\chi_1},\ e^{i\chi_2},\ 1),
\end{array}
\label{eq:VMNS}
\end{equation}
where $U_{MNS}$ is specified by the convention of Eq. (\ref{eq:MNS}), and 
$\Phi$ is a diagonal phase matrix involving the two Majorana phases,
$\chi_1$ and $\chi_2$.

Note that if the right-handed Majorana mass matrix had the structure $M_R 
\propto M^T_N M_N$, by the seesaw formula in Eq. (\ref{eq:seesaw}),
$M_{\nu_L}$ would be proportional to the identity matrix with three-fold  
degeneracy.  This suggests that some satisfactory $M'_R$ may be found leading
to an inverted neutrino mass hierarchy.  We now determine the corresponding
matrices which yield not only the correct mass spacing with an inverted 
hierarchy, but also the correct MNS mixing matrix.

For this purpose, let us choose the heaviest light neutrino mass $m'_2$ to 
be assigned some value to be specified later.  The other light neutrino masses
can then be determined from the observed mass squared differences in Eq. 
(\ref{eq:mix}) and $m'_2$ according to 
\begin{equation}
	m'_1 = \sqrt{m'^2_2 - \Delta m^2_{21}},\quad 
	m'_3 = \sqrt{m'^2_2 - |\Delta m^2_{32}|}.
\label{eq:mnupmasses}
\end{equation}
We invert the counterpart of Eq. (\ref{eq:mnudiag}) written for the 
inverted hierarchy case and identify the resultant $M'_\nu$ with the seesaw
formula involving the same Dirac neutrino matrix but the new right-handed 
Majorana matrix, $M'_R$:
\begin{equation}
\begin{array}{rcl}
	M'_\nu &=& U^*_{\nu_L} {\rm diag}(m'_1,\ m'_2,\ m'_3)U^\dagger_{\nu_L} 
		\\[0.1in]
		&=& - M_N M'^{-1}_R M^T_N.
\end{array}
\label{eq:mnup}
\end{equation}
By using the same unitary tranformation $U_{\nu_L}$ for the inverted hierarchy
case as in the normal hierarchy case and the masses $m'_i$ with 
$m'_3 < m'_1 < m'_2$ ordered in the diagonal matrix as above, we guarantee that 
the same desired MNS mixing matrix, $U_{MNS}$, and Majorana phase matrix,
$\Phi$, are obtained in the inverted case.  In general one will find that the 
mild normal hierarchy for the light neutrino mass matrix, $M_\nu$, is replaced 
by a completely different texture with no simple flavor symmetry obvious 
for the inverted hierarchy version.  The new right-handed Majorana matrix which 
accomplishes this follows from the seesaw formula in Eq. (\ref{eq:mnup}) by 
inversion:
\begin{equation}
	M'_R = - M^T_N M'^{-1}_\nu M_N.
\label{eq:mrp}
\end{equation}

The procedure described above for determining the new light neutrino mass matrix
$M'_\nu$ which obtains in the inverted hierarchy case, given the original
normal hierarchy model, can be applied for any grand unified model with a type~I
seesaw mechanism. It ensures
that the neutrino oscillation data for the mixing matrix will again be 
satisfactorily fit but now with an inverted neutrino mass spectrum.  Of crucial
importance is the fact that the Dirac neutrino mass matrix is the same for
both hierarchies,
for with a family unification group such as $SO(10)$, this mass matrix is 
closely related to that for the up quark sector, since the same set of Higgs 
representations are involved.  This means that the two different hierarchies 
arise solely from the different right-handed Majorana mass matrix textures.  
Hence the heavy right-handed neutrino spectra will differ, 
as will the possibilities for successful leptogenesis.  

The nature of the light neutrino spectrum of course has a direct bearing on 
the possible observation of neutrino-less double beta decay.  The relevant 
parameter is the effective mass which can be written as 
\begin{equation}
	\langle m_{ee} \rangle = |\sum_j m_j \left( U_{MNS}\Phi\right)^2_{1j}|.
\label{eq:ee}
\end{equation}
The same relation applies for both the normal and inverted hierarchies, with 
$m_j$ for the normal hierarchy replaced by the primed counterparts, $m'_j$, 
in the inverted case, with $j = 1,2,3$ taken in the same order. 

To make the distinctions more obvious, we shall first use as an example
a much studied $SO(10)$ grand unified model which 
explains well the quark mass and mixing data, as well as the lepton mass and 
mixing data with a natural normal neutrino mass hierarchy \cite{ab}.  
Based on the $SO(10)$ family symmetry with a $U(1) \times Z_2 \times Z_2$ 
flavor symmetry, one adjoint ${\bf 45}_H$, two pairs of ${\bf 16}_H$, 
$\overline{\bf 16}_H$, along with several Higgs fields in the ${\bf 10}_H$ 
and singlet representations, the nine quark and charged lepton masses and four 
CKM quark mixing angles and CP phase are well fitted with eight model parameters.  
The lopsided nature of the down quark and charged lepton mass matrices,
arising from an electroweak symmetry-breaking Higgs field in the ${\bf 16}_H$ 
representation, readily explains the small $V_{cb}$ quark mixing and near maximal 
$U_{\mu 3}$ atmospheric neutrino mixing for any reasonable right-handed Majorana 
neutrino mass matrix, $M_R$ \cite{abb}.  The type of solar neutrino solution is 
found to be controlled mainly by $M_R$ in this model.  In fact, the uncertain 
nature of this $M_R$ matrix has closely paralleled the uncertainty in the 
neutrino mixing data mentioned in the introduction.  When the model was initially 
proposed, a very simple form of $M_R$ led directly to the small mixing angle 
(SMA) solar neutrino solution favored by experiment at that time.  A more 
complicated structure with zero subdeterminant for the $2-3$ sector was later 
realized to obtain the large mixing angle (LMA) solar neutrino solution 
\cite{abLMA}.  The near degeneracy of the two lightest right-handed Majorana 
neutrinos and the resulting possibility of successful resonant leptogenesis was 
recognized in \cite{abres}.  

We shall simply begin with the relevant mass matrices and refer the interested 
reader to the literature cited for more specific details of this model.  
The two Dirac mass matrices for the charged leptons and neutrinos are given by
\begin{equation}
\begin{array}{ll}
M_N = \left(\matrix{ \eta & \delta_N & \delta'_N \cr \delta_N & 0 & \epsilon \cr
        \delta'_N & - \epsilon & 1\cr} \right)m_U,\
  & M_L = \left(\matrix{ 0 & \delta & \delta' e^{i \phi} \cr
  \delta & 0 & \sigma + \epsilon \cr \delta' e^{i\phi} &
  - \epsilon & 1\cr} \right)m_D.\\
\end{array}
\label{eq:Dmatrices}
\end{equation}
Here the Dirac matrices are written with the convention that the left-handed 
fields label the rows and the left-handed conjugate fields label the columns.
All nine quark and charged lepton masses, plus the three CKM angles and CP 
phase, are well fitted with the eight input parameters, 
\begin{equation}
\begin{array}{rlrl}
        m_U&\simeq 113\ {\rm GeV},&\qquad m_D&\simeq 1\ {\rm GeV},\\
        \sigma&= 1.83,&\qquad \epsilon&=0.147,\\
        \delta&= 0.00946,&\qquad \delta'&= 0.00827,\\
        \phi&= 119.4^\circ,&\qquad \eta&= 6 \times 10^{-6},\\
\end{array}
\label{eq:inputparam}
\end{equation}
defined at the GUT scale to fit the low scale observables after evolution 
downward from $\Lambda_{GUT}$. As initially proposed, the two parameters 
$\delta_N$ and $\delta'_N$ were set equal to zero, but we shall allow them
to assume also non-zero, but very small, values which can lead to a more nearly
satisfactory result for leptogenesis \cite{ablepto}.  Such values considered are
small enough so as not to destroy the good fit with experiment obtained 
initially. 

The most general form for the right-handed Majorana mass matrix considered in 
\cite{abLMA} which gives the large mixing angle (LMA) solar neutrino solution is
\begin{equation}
          M_R = \left(\matrix{c^2 \eta^2 & -b\epsilon\eta & a\eta\cr
                -b\epsilon\eta & \epsilon^2 & -\epsilon\cr
                a\eta & -\epsilon & 1\cr}\right)\Lambda_R,\\
\label{eq:MR}
\end{equation}
as explained by the structure of the Froggatt-Nielsen diagrams \cite{fn}.
The same parameter, $\epsilon$, appears as in Eqs. (\ref{eq:Dmatrices}) and
(\ref{eq:inputparam}) under the assumption of a universal coupling of the 
${\bf 45}_H$ Higgs field to all the matter fields.  For the parameters 
originally chosen in the model, this matrix exhibits a normal hierarchy which 
departs from a simple geometrical form when not all three of $a,\ b$ and $c$ 
are equal.  For the simple case where $\delta_N = \delta'_N = 0$, the seesaw 
formula of Eq. (\ref{eq:seesaw}) determines the light neutrino mass matrix to be
\begin{equation}
\begin{array}{rcl}
	M_\nu &=& - M_N M^{-1}_R M^T_N\\[0.1in] 
		&=& -\left(\matrix{ 0 & 
                        \frac{1}{a-b} \epsilon & 0\cr 
                        \frac{1}{a-b} \epsilon & \frac{b^2-c^2}{(a-b)^2} 
                        \epsilon^2 
                        & \frac{b}{b-a} \epsilon\cr 0 & \frac{b}{b-a} \epsilon 
                        & 1\cr} \right)m^2_U/\Lambda_R.
\end{array}
\label{eq:Mnu}
\end{equation}
The rather extreme hierarchy of $M_R$ is more than balanced by the 
normal hierarchy of $M_N$ and $M^T_N$ to yield a rather mild normal hierarchy
for $M_\nu$ for selected values of $a,\ b$ and~$c$.

We proceed to illustrate the connection between the inverted and normal 
versions of the model with a numerical example.  Let us choose the following
parameters which lead to a normal hierarchy,
\begin{equation}
\begin{array}{ll}
	a = 0.25 + 0.15i,\quad & b = 1.20 + 0.90i,\\[0.1in] 
	c = 0.25 + 0.25i,\quad & \Lambda_R = 2.90 \times 10^{14}\ {\rm GeV},
		\\[0.1in]
	\delta_N = -0.65 \times 10^{-5},\quad & \delta'_N = 
		-1.0 \times 10^{-5},\\
\label{eq:mrparam}
\end{array}
\end{equation}
and were recently found to yield nearly successful baryogenesis with a resonant 
leptogenesis mechanism \cite{ablepto}.  

With this choice of parameters listed, the right-handed Majorana matrix is 
given by  
\begin{equation}
\begin{array}{rcl}
	M_R &=& \left(\matrix{4.5i \times 10^{-12} & -(1.058+0.794i) \times 10^{-6} 
		& (1.5 + 0.9i)\times 10^{-6}\cr
		-(1.058+0.794i) \times 10^{-6} & 0.0216 & -0.147\cr
		(1.5 + 0.9i) \times 10^{-6} & -0.147 & 1\cr}\right)\\[0.3in]
	      & & \times\ 2.9 \times 10^{14}\ {\rm GeV},
\end{array}
\label{eq:mr}
\end{equation}
from which the seesaw formula of Eq. (\ref{eq:seesaw}) yields
\begin{equation}
\begin{array}{rcl}
	M_\nu &=& -\left(\matrix{(7.03-5.55i) \times 10^{-5} & 
		-0.0954 + 0.0753i & -(8.16-4.79i) \times 10^{-5}\cr
		-0.0954 + 0.0753i & 0.238-0.165i & 0.341 - 0.130i\cr
		-(8.16-4.79i) \times 10^{-5} & 0.341 - 0.130i & 1\cr}\right)\\[0.3in]
	      & & \times\ 0.0440\ {\rm eV}.
\end{array}
\label{eq:mnu}
\end{equation}
Some results obtained in the model for this normal hierarchy version are 
\begin{equation}
\begin{array}{rl}
	&M_3 = 2.96 \times 10^{14},\ M_2 \simeq M_1 = 3.06 \times 
		10^8\ {\rm GeV},\\[0.1in]
	&m_3 = 51.0,\ \quad m_2 = 8.8,\ 
		\quad m_1 = 2.8\ {\rm meV},\\[0.1in]
	&\langle m_{ee} \rangle = 0.44\ {\rm meV},\quad \eta_B = 2.2 \times 
		10^{-10},
\end{array}
\end{equation}
for the heavy right-handed and light left-handed neutrino masses, the effective
neutrino-less double beta decay mass $\langle m_{ee}\rangle$, and the net baryon
number to photon number asymmetry ratio for the Universe, $\eta_B$.  The very
small departures of $\delta_N$ and $\delta'_N$ from zero serve to make the 
lighter two heavy right-handed neutrinos more nearly degenerate and thus enhance
the resonant leptogenesis effect.  We refer the reader to reference \cite{abres}
for the explicit neutrino mixing angle results which are in excellent agreement
with the present data.

To obtain a corresponding model with inverse hierarchy, we select 
$m'_2 = 0.10$ eV, which lies below the maximum of 0.23 eV allowed @ 95\% c.l. 
for nearly degenerate neutrinos by the recent WMAP data in combination with 
the 2dF Galaxy Redshift Survey and Lyman-$\alpha$ forest power spectrum 
\cite{WMAP}.  The procedure elaborated above in Eqs. (\ref{eq:mnupmasses}) 
and (\ref{eq:mnup}) with the help of the unitary transformation matrix, 
\begin{equation}
	U_{\nu_L} = \left(\matrix{0.799 + 0.353i & -0.198 + 0.443i & 
			-0.0134 + 0.0391i\cr
		0.446 + 0.115i & 0.196 - 0.772i & 0.130 - 0.370i\cr
		-0.157 + 0.020i & 0.028 + 0.360i & 0.002 - 0.918i\cr}\right)
\end{equation}
determined from the normal hierarchy case as in Eq. (\ref{eq:mnudiag}),
then leads to the light neutrino mass matrix, 
\begin{equation}
\begin{array}{rcl}
	M'_\nu &=& -\left(\matrix{-0.416+0.467i & -0.744+0.604i &
			0.317-0.109i\cr
		-0.744+0.604i & 0.565-0.346i & 0.098-0.078i\cr
		0.317-0.109i & 0.098-0.078i & 1\cr}\right)\\[0.3in]  
	     & & \times\ 0.0836\ {\rm eV},
\end{array}
\label{eq:mpnu}
\end{equation}
in the inverted hierarchy version considered, from which the required $M'_R$ is 
determined to be 
\begin{equation}
\begin{array}{rcl}
	M'_R &=& \left(\matrix{(1.397+0.318i) \times 10^{-10} & 
		(1.340-0.025i) \times 10^{-6} & -(1.028+0.058i) \times 10^{-5}\cr
		(1.340-0.025i) \times 10^{-6} & 0.0204-0.0006i & -0.142+0.003i\cr
		-(1.028+0.058i) \times 10^{-5} & -0.142+0.003i & 1\cr}\right)\\[0.3in]
	      & & \times\ 1.46 \times 10^{14}\ {\rm GeV}.
\end{array}
\label{eq:mpr}
\end{equation}
From Eq. (\ref{eq:mnupmasses}) and these matrices we find 
\begin{equation}
\begin{array}{rl}
	&m'_2 = 0.1000,\ m'_1 = 0.0996,\ m'_3 = 0.0866\ {\rm eV},\\[0.1in]
	&M'_3 = 1.49 \times 10^{14},\ \ M'_2 = 3.10 \times 10^{10},\ 
		M'_1 = 8.77 \times 10^3\ {\rm GeV},\\[0.1in]
	&\langle m'_{ee} \rangle = 0.044\ {\rm eV},\quad \eta'_B = 8.5 \times 
		10^{-22}.
\end{array}
\end{equation}

While $M_R$ and $M'_R$ do not appear to be significantly different, the opposite
signs of the 12, 21 and 13, 31 elements for the two matrices play an important
role.  Moreover, the 2-3 sector subdeterminant fails to vanish in the latter case.
This accounts for the huge hierarchy in the right-handed neutrinos in the inverted 
case, which departs from the near degeneracy of the two lighter right-handed 
neutrinos in the normal case.  As a result, the nearly successful resonant 
leptogenesis solution with the normal hierarchy is completely lost. The heavy 
right-handed neutrinos in the inverted case span such a huge hierarchy that the 
lightest is much lighter than the $10^9 - 10^{10}$ GeV required for successful 
leptogenesis without a resonant enhancement \cite{lepto}.

More striking are the differences in the light left-handed neutrino sector, 
where the mass spectrum is indeed inverted but nearly degenerate by the choice of 
$m'_2$, as illustrated in Fig.~1.  While the normal hierarchy $M_\nu$ matrix has 
two near texture zeros in the 11, 13 and 31 elements, the elements of $M'_\nu$ are 
all rather comparable in magnitude.  This is arises since the determinant of 
$M'_\nu$ is three orders of magnitude larger than that for $M_\nu$, i.e., 
$|m'_3 m'_2 m'_1| \sim 700 |m_3 m_2 m_1|$.  Such an inverted spectrum is highly 
unstable against very small changes in the matrix elements of $M'_R$.  
It is also very unstable against radiative corrections upon evolution downward 
from the GUT scale, since two of the nearly degenerate masses have approximately 
the same CP parity \cite{radcorr}.  Mainly for these reasons, we consider 
the inverted hierarchy solution in this model to be much less satisfactory than 
the normal hierarchy solution.  

We regard this result to be a general feature of grand unified models with a 
type I seesaw mechanism.  Unfortunately there are very few models in the literature
as predictive as this with which we are prepared to test this prediction.  However,
we can further illustrate the claim by considering the whole class of models for 
which the charged lepton mass matrix is diagonal in flavor space.  Recall that for
the specific model examined in detail above, the charged lepton mass matrix is
lopsided in that basis.  

With the charged lepton mass matrix, $M_L$, diagonal in the flavor basis, the 
unitary transformation, $U_{L_L}$, is just the identity matrix.  Successful models
in this class then require that $U_{\nu_L} \simeq U_{MNS}$ by Eq. (\ref{eq:VMNS}),
where for our purposes we can drop the two Majorana phases without loss of 
generality.  We set $U_{MNS}$ equal to that in Eq. (\ref{eq:MNS}) with the 
approximations $c_{23} \simeq s_{23} \simeq 1/\sqrt{2}$,
$s_{12} \simeq \sqrt{2/7},\ c_{12} \simeq \sqrt{5/7}$ and $s_{13} = 0$.  By
selecting the normal hierarchy neutrino masses found earlier, and inverting 
Eq. (\ref{eq:mnudiag}) we find for the light neutrino mass matrix,
\begin{equation}
\begin{array}{rcl}
	M_\nu &=& U^*_{\nu_L}{\rm diag}(0.0028,\ 0.0088,\ 0.051)U^\dagger_{\nu_L}
		\\[0.1in]
		&=& \left(\matrix{0.451 & 0.192 & -0.192\cr 
			0.192 & 2.904 & 2.196\cr -0.192 & 2.196 & 2.904\cr}\right)
			\times 0.01\ {\rm eV.}\cr
\end{array}
\label{toymnu}
\end{equation}
This matrix clearly has the texture for a reasonably stable normal mass hierarchy.

To investigate a corresponding inverted mass hierarchy version, we again select
$m'_2 = 0.10$ eV and find with the same $U_{\nu_L} = U_{MNS}$ the new light
neutrino mass matrix, 
\begin{equation}
\begin{array}{rcl}
	M'_\nu &=& U^*_{\nu_L}{\rm diag}(0.0996,\ 0.10,\ 0.0866)U^\dagger_{\nu_L}
			\\[0.1in]
		&=& \left(\matrix{9.975 & 0.011 & -0.011\cr
			0.011 & 9.325 & -0.665\cr -0.011 & -0.665 & 9.325\cr}\right)
			\times 0.01\ {\rm eV}.
\end{array}
\label{toymnup}
\end{equation}
Note that if the small off-diagonal elements of this matrix were neglected, the 
neutrino mass hierarchy would be normal with the lowest two levels degenerate.
It is only the small 23 and 32 off-diagonal corrections which convert the 
hierarchy into an inverted one.  This matrix is thus much more sensitive to 
small changes in the underlying right-handed Majorana matrix structure for each
model in the class considered.  On the other hand, the dominant diagonal 
elements set the scale for the determinant of the matrix and thus the product of 
the three neutrino masses, which again is about 700 times larger than that for 
$M_\nu$.  

These results should be contrasted with those for models involving a softly 
broken $L_e - L_\mu - L_\tau$ flavor symmetry.  There the light neutrino 
mass matrix has large 12, 13, 21 and 31 matrix elements with small values
for the other elements.  Hence the results are less sensitive to 
instabilities resulting from small perturbations away from the desired 
entries, and in such models no evolution from a high mass scale is typically 
involved.

In summary, we have shown that to every successful normal neutrino mass 
hierarchy solution of a grand unified model corresponds an inverted hierarchy 
solution with exactly the same MNS mixing matrix.  The inverted left-handed 
neutrino spectrum illustrated is nearly degenerate and is highly unstable to 
small changes in the parameters for the right-handed Majorana mass matrix as 
well as to radiative corrections upon evolution from the grand unified scale.  
This suggests that future observation of an inverted 
hierarchy would tend to disfavor grand unified models based on the conventional 
type I seesaw mechanism.  On the other hand, grand unified models with type~II
or type III seesaw mechanisms, where a left-handed Majorana mass matrix can 
arise from direct or induced Higgs triplet contributions, or models based on a 
conserved $L_e - L_\mu - L_\tau$ lepton 
number would then be favored.  Conversely, observation of a normal hierarchy
would eliminate the conserved $L_e - L_\mu - L_\tau$ number models in favor
of the grand unified models.  Successful leptogenesis in the latter type models 
is also more favorable.\\[0.3in]

The author thanks Boris Kayser for urging him to sharpen the argument
that grand unified models with a type I seesaw mechanism tend to favor a 
normal light neutrino mass spectrum.  He also acknowledges suggestions made by
Stephen Barr.  The author thanks the Theory Group at Fermilab for its kind
hospitality.  Fermilab is operated by Universities Research Association Inc.
under contract No. DE-AC02-76CH03000 with the Department of Energy.

%
%

\newpage
%
\begin{figure}
\vspace{-2in}
\center{
\epsfbox{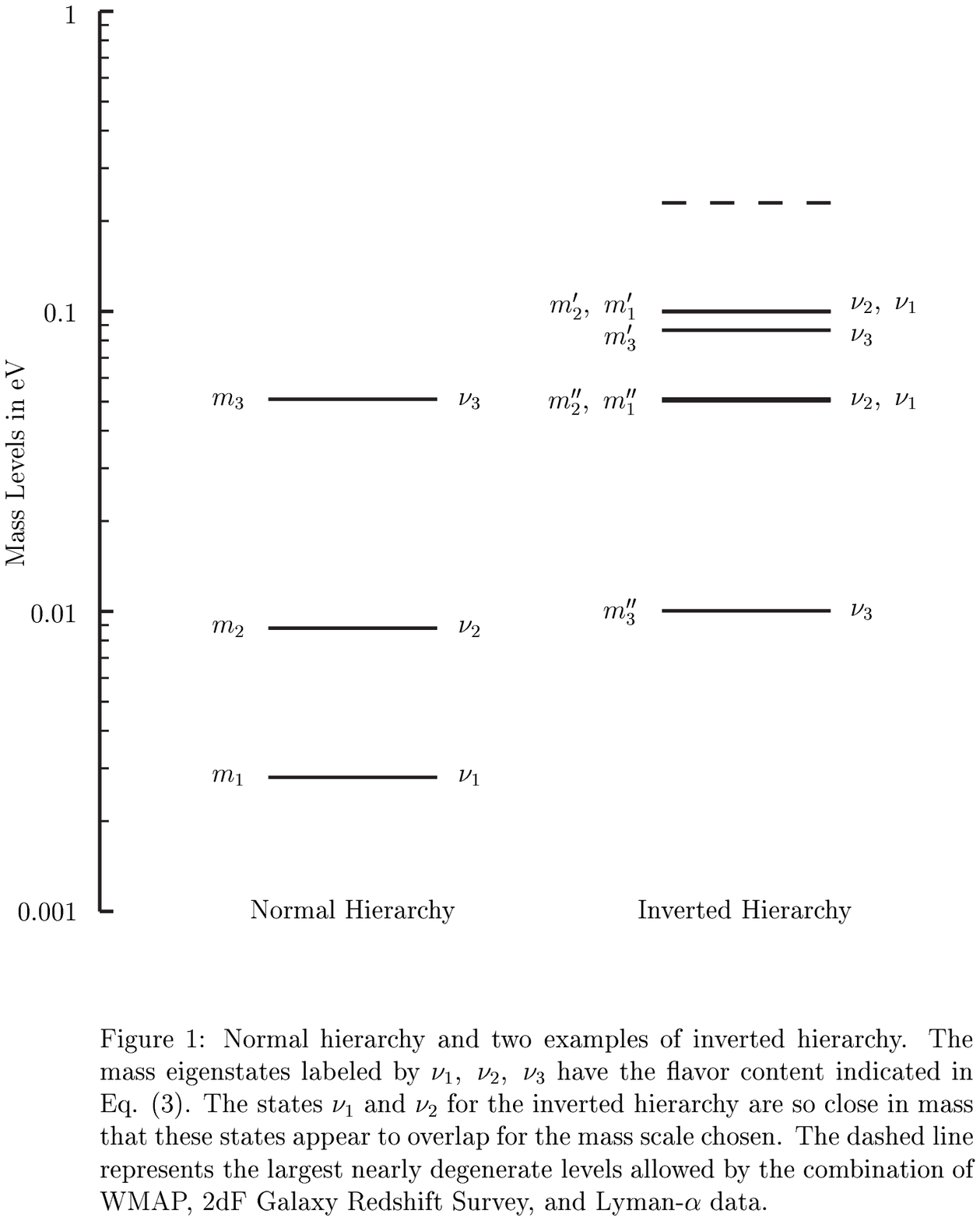}
}
\vspace {0in}
\end{figure}

\end{document}